\documentclass[11pt]{article}

\usepackage{amssymb}
\usepackage{cite}
\usepackage{amsmath,mathrsfs}
\usepackage{url}

\def\pa{\partial}

\def\ka{\kappa}
\def\al{\alpha}

\def\ii{\textrm i}
\def\ee{\textrm e}

\def\ud{\textrm{d}}
\def\bn{{\boldsymbol \nabla}}

\newcommand{\be}{\begin{equation}}
\newcommand{\en}{\end{equation}}
\newcommand{\bea}{\begin{eqnarray}}
\newcommand{\ena}{\end{eqnarray}}

\begin{document}

\vspace*{1.0cm}
\noindent
{\bf
{\large
\begin{center}
Towards a novel approach to semi-classical gravity
\end{center}
}
}

\vspace*{.5cm}
\begin{center}
Ward Struyve\\
Department of Physics, University of Liege\\
All\'ee du 6 Ao\^ut 10, 4000 Liege, Belgium. \\
E--mail: ward.struyve@ulg.ac.be
\end{center}

\begin{abstract}
\noindent
Semi-classical gravity is an approximation to quantum gravity where gravity is treated classically and matter quantum mechanically. Matter is described by quantum field theory on curved space-time, whereas gravity is described by a space-time metric which satisfies Einstein's field equations. In the usual approach to semi-classical gravity, the matter source term in Einstein's field equations is given by the expectation value of the energy-momentum tensor operator. In this paper, we suggest an alternative approach based on Bohmian mechanics. In Bohmian mechanics, a quantum system is described by an actual configuration (e.g., an actual scalar field), which evolves under the influence of the wave function. The idea is to consider the actual energy-momentum tensor corresponding to this configuration as the matter source term in Einstein's field equations. This approach is expected to improve upon the usual approach.
\end{abstract}

\renewcommand{\baselinestretch}{1.1}
\bibliographystyle{unsrt}
\bibliographystyle{plain}

\section{Introduction}
Quantum gravity is often considered to be the holy grail of theoretical physics. One approach is canonical quantum gravity, which concerns the Wheeler-DeWitt equation and which is obtained by applying the usual quantization methods (which were so successful in the case of high energy physics) to Einstein's field equations. However, this approach suffers from a host of problems, some of technical and some of conceptual nature (such as finding solutions to the Wheeler-DeWitt equation, the problem of time, \dots). For this reason one often resorts to a semi-classical approximation where gravity is treated classically and matter quantum mechanically \cite{wald94,kiefer04}. The hope is that such an approximation is easier to analyse and yet reveals some effects of quantum gravitational nature.

In the usual approach to semi-classical gravity, matter is described by quantum field theory on curved space-time. For example, in the case the matter is described by a quantized scalar field, the state vector can be considered to be a functional $\Psi(\phi)$ on the space of fields, which satisfies a particular Schr\"odinger equation
\be
\ii \pa_t \Psi(\phi,t) = {\widehat H}(\phi,g) \Psi(\phi,t) \,,
\label{0.001}
\en
where the Hamiltonian operator ${\widehat H}$ depends on the space-time metric $g$. This metric satisfies Einstein's field equations
\be
G_{\mu \nu} (g) = 8\pi G \langle \Psi | {\widehat T}_{\mu \nu} (\phi,g) |\Psi\rangle \,,
\label{0.002}
\en
where the source term is given by the expectation value of the energy-momentum tensor operator.

This semi-classical approximation of course has limited validity. For example, it will form a good approximation when the matter state approximately corresponds to a classical state (i.e., a coherent state), but will fail to be so when the state is a macroscopic superposition of such states. Namely, for such a superposition $\Psi = (\Psi_1 + \Psi_2)/{\sqrt 2}$, we have that $\langle \Psi| {\widehat T}_{\mu \nu} |\Psi\rangle \approx \left( \langle \Psi_1| {\widehat T}_{\mu \nu} |\Psi_1\rangle + \langle \Psi_2| {\widehat T}_{\mu \nu} |\Psi_2\rangle \right)/2$, so that the gravitational field is affected by two matter sources, one coming from each term in the superposition. However, one expects that according to a full theory for quantum gravity, the states $|\Psi_1\rangle$ and $|\Psi_2\rangle$ each have their own gravitational field and that the total state is a superposition of those. And, indeed, Page and Geilker showed with an experiment that this semi-classical theory is not adequate \cite{page81,kiefer04}. 

Of course, as already noted by Page and Geilker, it could be that this problem is not due to fact gravity is treated classically, but due to the choice of the version of quantum theory. Namely, Page and Geilker adopted the Many Worlds point of view, according to which the wave function never collapses. However, according to standard quantum theory the wave function is supposed to collapse during a measurement. Which physical processes act as measurements is of course rather vague and is the source of the measurement problem. But it could be that such collapses explain the outcome of their experiment. If an explanation of this type is sought, one should consider so-called spontaneous collapse theories, where collapses are objective, random processes that do not in a fundamental way depend on the notion of measurement. (See \cite{diez-tjedor12} and \cite{derakhshani14} for actual proposals combining such a spontaneous collapse approach with respectively \eqref{0.002} and its non-relativistic version.)

In this paper, we consider an alternative to standard quantum mechanics, called Bohmian mechanics \cite{bohm93,holland93b,duerr09,duerr12}. Bohmian mechanics solves the measurement problem by introducing an actual configuration (particle positions in the non-relativistic domain, particle positions or fields in the relativistic domain \cite{struyve11}) that evolves under the influence of the wave function. According to this approach, instead of coupling classical gravity to the wave function, it is natural to couple it to the actual matter configuration. For example, in the case of a scalar field there is an actual field $\phi_{B}$ whose time evolution is determined by the wave functional $\Psi$. There is an energy-momentum tensor $T_{\mu \nu}(\phi_{B},g)$ corresponding to this scalar field and this tensor can be introduced as the source term in Einstein's field equations:
\be
G_{\mu \nu} (g) = T_{\mu \nu}(\phi_{B},g) \,.
\label{0.003}
\en
This approach immediately solves the problem with the macroscopic superposition, since the energy-momentum tensor will correspond to just one of the macroscopic matter distributions. 

However, there is an immediate problem with this approach, namely that equation \eqref{0.003} is not consistent. The Einstein tensor $G_{\mu \nu}$ is identically conserved, i.e., $\nabla^\mu G_{\mu \nu} \equiv 0$. So the Bohmian energy-momentum tensor $T_{\mu \nu}(\phi_{B},g)$ must be conserved as well. However, the equation of motion for the scalar field does not guarantee this. (Similarly, in the Bohmian approach to non-relativistic systems, the energy is generically not conserved.) 

As explained in \cite{struyve15}, the root of the problem seems to be the gauge invariance, which in this case is the invariance under spatial diffeomorphisms. Because the scalar field and the space-time metric are connected by spatial diffeomorphisms, it seems that one can not just assume the metric to be classical without also assuming the scalar field $\phi_{B}$ to be classical (in which case the energy-momentum tensor is conserved).

A similar problem arises when we consider a Bohmian semi-classical approximation to scalar electrodynamics, which describes a scalar field interacting with an electromagnetic field. In this case, the wave equation for the scalar field is of the form
\be
\ii \pa_t \Psi(\phi,t) = {\widehat H}(\phi,A) \Psi(\phi,t) \,,
\en
where $A$ is the vector potential. There is also a Bohmian scalar field $\phi_B$ and a charge current $ j^\nu(\phi_B,A)$ that could act as the source term in Maxwell's equations
\be
\pa_\mu F^{\mu \nu}(A) = j^\nu(\phi_B,A)  \,,
\en
where $F^{\mu \nu}$ is the electromagnetic field tensor. In this case, we have $\pa_\nu \pa_\mu F^{\mu \nu} \equiv 0$ due to the anti-symmetry of $F^{\mu \nu}$. As such, the charge current must be conserved. However, the Bohmian equation of motion for the scalar field does not imply conservation. Hence, just as in the case of gravity, a consistency problem arises. As explained in \cite{struyve15}, this problem can be overcome by eliminating the gauge invariance, either by assuming some gauge fixing or (equivalently) by working with gauge-independent degrees of freedom. In this way, we can straightforwardly derive a semi-classical approximation starting from the full Bohmian approach. For example, in the Coulomb gauge, the result is that there is an extra current $j^\nu_Q$ which appears in addition to the usual charge current and which depends on the quantum potential, so that Maxwell's equations read 
\be
\pa_\mu F^{\mu \nu}(A) = j^\nu(\phi_B,A) +  j^\nu_Q(\phi_B,A) \,.
\en

While it is easy to eliminate the gauge invariance in the case of electrodynamics, this is notoriously difficult in the case of general relativity. One can formulate a Bohmian approach for the Wheeler-DeWitt equation for a scalar matter field interacting with gravity, but the usual formulation does not explicitly eliminate the gauge freedom arising from spatial diffeomorphism invariance. Our expectation is that one could find a semi-classical approximation given such a formulation. At least we find our expectation confirmed in simplified models, called mini-superspace models, where this invariance is eliminated. We will illustrate this for the model described by the homogeneous and isotropic Friedmann-Lema\^itre-Robertson-Walker metric and a uniform scalar field. 

In this paper, we are merely concerned with the formulation of Bohmian semi-classical approximations. Practical applications will be studied elsewhere. Such applications already have been studied in the context of quantum chemistry, see section \ref{scanrqm}. It appears that Bohmian semi-classical approximations yield better or equivalent results compared to the usual semi-classical approximation. (They are better in the sense that they are closer to the exact quantum results). This provides good hope that also in other contexts, such as quantum gravity, the Bohmian approach also gives better results. Potential applications might be found in inflation theory, where the back-reaction from the quantum fluctuations onto the classical background can be studied, or in black hole physics, to study the back-reaction from the Hawking radiation onto space-time.

The outline of the paper is as follows. After introducing Bohmian mechanics in section \ref{bm}, we will discuss how to derive a Bohmian semi-classical approximation in the context of non-relativistic quantum mechanics. Semi-classical approximations to other quantum theories can be derived in a similar way. We present such approximations for scalar quantum electrodynamics in section \ref{sqed} and for a mini-superspace model in section \ref{mini}. More examples and details can be found in \cite{struyve15}.

\section{Bohmian mechanics}\label{bm} 
\subsection{Non-relativistic quantum mechanics}
Non-relativistic Bohmian mechanics (also called pilot-wave theory or de Broglie-Bohm theory) is a theory about point-particles in physical space moving under the influence of the wave function \cite{bohm93,holland93b,duerr09,duerr12}. The equation of motion for the configuration $X=({\bf X}_1,\dots,{\bf X}_n)$ of the particles is given by{\footnote{Throughout the paper we assume units in which $\hbar=c=1$.}}
\be
{\dot X}(t) = v^\psi(X(t),t)\,,
\label{0.01}
\en
where $v^\psi=({\bf v}^\psi_1, \dots , {\bf v}^\psi_n)$, with
\be
{\bf v}^\psi_k = \frac{1}{m_k} {\textrm{Im}}\left( \frac{\boldsymbol{\nabla}_k \psi}{\psi} \right) =  \frac{1}{m_k} {\boldsymbol \nabla}_k S  
\label{0.02}
\en 
and $\psi = |\psi| \ee^{\ii S}$. The wave function $\psi(x,t)=\psi({\bf x}_1,\dots,{\bf x}_n)$ itself satisfies the non-relativistic Schr\"odinger equation
\be
\ii \pa_t \psi(x,t) = \left( - \sum^n_{k=1} \frac{1}{2m_k} \nabla^2_k + V(x) \right) \psi(x,t) \,.
\label{0.03}
\en

For an ensemble of systems all with the same wave function $\psi$, there is a distinguished distribution given by $|\psi|^2$, which is called the {\em quantum equilibrium distribution}. This distribution is {\em equivariant}. That is, it is preserved by the particles dynamics \eqref{0.01} in the sense that if the particle distribution is given by $|\psi(x,t_0)|^2$ at some time $t_0$, then it is given by $|\psi(x,t)|^2$ at all times $t$. This follows from the fact that any distribution $\rho$ that is transported by the particle motion satisfies the continuity equation
\be
\pa_t \rho + \sum^n_{k=1} {\boldsymbol \nabla}_k \cdot ({\bf v}^\psi_k \rho) = 0 
\label{0.04}
\en
and that $|\psi|^2$ satisfies the same equation, i.e.,
\be
\pa_t |\psi|^2 + \sum^n_{k=1} {\boldsymbol \nabla}_k \cdot ({\bf v}^\psi_k |\psi|^2) = 0 \,,
\label{0.041}
\en
as a consequence of the Schr\"odinger equation. It can be shown that for a typical initial configuration of the universe, the (empirical) particle distribution for an actual ensemble of subsystems within the universe will be given by the quantum equilibrium distribution \cite{duerr92a,duerr09,duerr12}. Therefore for such a configuration Bohmian mechanics reproduces the standard quantum predictions. 

Note that the velocity field is of the form $j^\psi/|\psi|^2$, where $j^\psi=({\bf j}^\psi_1,\dots,{\bf j}^\psi_n)$ with ${\bf j}^\psi_k=  {\textrm{Im}}( \psi^* \boldsymbol{\nabla}_k \psi )/m_k $ is the usual quantum current. In other quantum theories, such as for example quantum field theories, the velocity can be defined in a similar way by dividing the appropriate current by the density. In this way equivariance of the density will be ensured. (See \cite{struyve09a} for a treatment of arbitrary Hamiltonians.)

This theory solves the measurement problem. Notions such as measurement or observer play no fundamental role. Instead measurement can be treated as any other physical process.  

There are two aspects of the theory that are important for deriving the semi-classical approximation. Firstly, Bohmian mechanics allows for an unambiguous analysis of the classical limit. Namely, the classical limit is obtained whenever the particles (or at least the relevant macroscopic variables, such as the center of mass) move classically, i.e., satisfy Newton's equation. By taking the time derivative of \eqref{0.01}, we find that
\be
m_k {\ddot {\bf X}}_k(t) = -{\boldsymbol{\nabla}}_k (V(x)+Q^\psi(x,t))\big|_{x=X(t)}\,,
\label{0.07}
\en
where 
\be
Q^\psi = -\sum^n_{k=1}\frac{1}{2m_k}\frac{\nabla^2_k |\psi|}{|\psi|}
\label{0.08}
\en
is the quantum potential. Hence, if the quantum force $-{\boldsymbol{\nabla}}_kQ^\psi$ is negligible compared to the classical force $-{\boldsymbol{\nabla}}_kV$, then the $k$-th particle approximately moves along a classical trajectory. 

Another aspect of the theory is that it allows for a simple and natural definition for the wave function of a subsystem \cite{duerr92a,duerr09}. Namely, consider a system with wave function $\psi(x,y)$ where $x$ is the configuration variable of the subsystem and $y$ is the configuration variable of its environment. The actual configuration is $(X,Y)$, where $X$ is the configuration of the subsystem and $Y$ is the configuration of the other particles. The wave function of the subsystem $\chi(x,t)$, called the {\em conditional wave function}, is then defined as
\be
\chi(x,t) = \psi(x,Y(t),t).
\label{0.05}
\en
This is a natural definition since the trajectory $X(t)$ of the subsystem satisfies 
\be
{\dot X}(t) = v^\psi(X(t),Y(t),t) = v^\chi(X(t),t) \,. 
\label{0.06}
\en
That is, for the evolution of the subsystem's configuration we can either consider the conditional wave function or the total wave function (keeping the initial positions fixed). (The conditional wave function is also the wave function that would be found by a natural operationalist method for defining the wave function of a quantum mechanical subsystem \cite{norsen14}.) The time evolution of the conditional wave function is completely determined by the time evolution of $\psi$ and that of $Y$. This makes that the conditional wave function does not necessarily satisfy a Schr\"odinger equation, although in many cases it does. This wave function collapses according to the usual text book rules when an actual measurement is performed.

\subsection{Quantum field theory}
We will also consider semi-classical approximations to quantum field theories. More specifically, we will consider bosonic quantum field theories. In Bohmian approaches to such theories it is most easy to introduce actual field variables rather than particle positions \cite{struyve10,struyve11}. To illustrate how this works, let us consider the free massless real scalar field (for the treatment of other bosonic field theories see \cite{struyve10}). Working in the functional Schr\"odinger picture, the quantum state vector is a wave functional $\Psi(\phi)$ defined on a space of scalar fields in 3-space and it satisfies the functional Schr\"odinger equation
\be
\ii \pa_t \Psi(\phi,t) = \frac{1}{2}\int d^3 x \left(- \frac{\delta^2}{\delta \phi({\bf x})^2} + {\boldsymbol \nabla} \phi({\bf x}) \cdot{\boldsymbol \nabla} \phi({\bf x}) \right) \Psi(\phi,t) \,.
\label{0.09}
\en
The associated continuity equation is
\be
\pa_t |\Psi(\phi,t)|^2 + \int d^3 x \frac{\delta}{\delta \phi({\bf x})} \left( \frac{\delta S (\phi,t)}{\delta \phi({\bf x})} |\Psi(\phi,t)|^2 \right) = 0 \,,
\label{0.10}
\en
where $\Psi = |\Psi| \ee^{\ii S}$. This suggests the guidance equation 
\be
\dot \phi ({\bf x},t) = \frac{\delta S(\phi,t) }{\delta \phi({\bf x})} \bigg|_{\phi({\bf x}) = \phi({\bf x}, t)} \,.
\label{0.11}
\en
(Note that in this case we did not distinguish notationally the actual field variable from the argument of the wave functional.) Taking the time derivative of this equation results in 
\be
\square \phi({\bf x},t) = - \frac{\delta Q^\Psi(\phi,t)}{\delta \phi({\bf x})}\bigg|_{\phi({\bf x}) = \phi({\bf x}, t)} \,,
\label{0.12}
\en
where
\be
Q^\Psi = - \frac{1}{2|\Psi|} \int d^3 x \frac{\delta^2 |\Psi|}{\delta \phi({\bf x})^2}
\label{0.13}
\en
is the quantum potential. The classical limit is obtained whenever the quantum force, i.e., the right hand side of equation \eqref{0.12}, is negligible. Then the field approximately satisfies the classical field equation $\square \phi=0$. 

One can also consider the conditional wave functional of a subsystem. A subsystem can in this case be regarded as a system confined to a certain region in space. The conditional wave functional for the field confined to that region is then obtained from the total wave functional by conditioning over the actual field value on the complement of that region. However, in the following we will not consider this kind of conditional wave functional. Rather, there will be other degrees of freedom, like for example other fields, which will be conditioned over.

This Bohmian approach is not Lorentz invariant. The guidance equation \eqref{0.11} is formulated with respect to a preferred reference frame and as such violates Lorentz invariance. This violation does not show up in the statistical predictions given quantum equilibrium, since the theory makes the same predictions as standard quantum theory which are Lorentz invariant.{\footnote{Actually, this statement needs some qualifications since regulators need to be introduced to make the theory and its statistical predictions well-defined \cite{struyve10}.}} The difficulty in finding a Lorentz invariant theory resides in the fact that any adequate formulation of quantum theory must be non-local \cite{goldstein11b}. One approach to make the Bohmian theory Lorentz invariant is by introducing a foliation which is determined by the wave function in a covariant way \cite{duerr14}. In this paper, we will not attempt to maintain Lorentz invariance. As such, the Bohmian semi-classical approximations will not be Lorentz invariant, (very likely) not even concerning the statistical predictions. This is in contrast with the usual approach like the one for gravity given by \eqref{0.001} and \eqref{0.002} which is fully Lorentz invariant. However, this does not take away the expectation that the Bohmian semi-classical approximation will give better or at least equivalent results compared to the usual approach. 

\subsection{Quantum gravity}\label{qg}
In canonical quantum gravity, the state vector is a functional $\Psi(h,\phi)$ on the space of 3-metrics $h_{ij}({\bf x})$ on a 3-dimensional manifold and fields $\phi({\bf x})$ (in the case the matter is described by a quantized scalar field). The wave functional is static and merely satisfies the constraints \cite{kiefer04}
\be
{\mathcal H} \Psi(h,\phi) = 0 \,,
\en
\be
{\mathcal H}_i \Psi(h,\phi) = 0 \,.
\en
Their explicit forms are not important here. The latter constraint expresses the fact that the wave functional is invariant under infinitesimal diffeomorphisms of 3-space. The former equation is the Wheeler-DeWitt equation. It is believed that this equation contains the dynamical content of the theory. However, it is as yet not clear how this dynamical content should be extracted. This is the problem of time \cite{kuchar92,kiefer04}.

In the Bohmian approach, there is an actual 3-metric and a scalar field, whose dynamics depends on the wave functional \cite{shtanov96,goldstein04,pinto-neto05}. The dynamics expresses how the Bohmian configuration changes along a succession of 3-dimensional space-like surfaces.{\footnote{The succession of the surfaces is determined by the lapse function and different choices of lapse function lead to different Bohmian dynamics. This implies that the dynamics is not invariant under space-time diffeomorphisms. Again, the root of the problem is the non-locality of quantum theory.}} Although the wave function is stationary, the Bohmian configuration will change along these surfaces for generic wave functions. This is how the Bohmian approach solves the problem of time. 

Some cosmological applications of the Bohmian approach to quantum gravity are the explanation of the quantum-to-classical transition in inflation theory \cite{hiley95,pinto-neto12a} and the study of space-time singularities \cite{pinto-neto12b,pinto-neto13,falciano15}.

\section{Non-relativistic quantum mechanics}\label{scanrqm}

\subsection{Usual versus Bohmian semi-classical approximation}
Consider a composite system of just two particles. The usual semi-classical approach (also called the mean-field approach) goes as follows. Particle 1 is described quantum mechanically, by a wave function $\chi({\bf x}_1,t)$, which satisfies the Schr\"odinger equation
\be
\ii \pa_t \chi({\bf x}_1,t) =  \left[ - \frac{1}{2m_1}\nabla^2_1  + V({\bf x}_1,{\bf X}_2(t)) \right] \chi({\bf x}_1,t) \,,
\label{1}
\en
where the potential is evaluated for the position of the second particle ${\bf X}_2$, which satisfies Newton's equation
\begin{align}
m_2 {\ddot {\bf X}}_2(t) &=   -\left\langle \chi \left| {\boldsymbol \nabla}_2   V({\bf x}_1,{\bf x}_2)  \big|_{{\bf x}_2={\bf X}_2(t)} \right| \chi \right\rangle \nonumber\\
&= \int d^3x_1|\chi({\bf x}_1,t)|^2  [-{\boldsymbol \nabla}_2 V({\bf x}_1,{\bf x}_2)] \Big|_{{\bf x}_2={\bf X}_2(t)}\,.
\label{2}
\end{align}
So the force on the right-hand-side is averaged over the quantum particle.

An alternative semi-classical approach based on Bohmian mechanics was proposed independently by Gindensperger {\em et al.}\ \cite{gindensperger00} and Prezhdo and Brooksby \cite{prezhdo01}. In this approach there is also an actual position for particle 1, denoted by ${\bf X}_1$, which satisfies the equation
\be
{\dot {\bf X}}_1(t) = {\bf v}^\chi({\bf X}_1(t),t) \,,
\label{3}
\en
where
\be
{\bf v}^\chi = \frac{1}{m_1} {\textrm{Im}} \frac{\bn \chi}{\chi} \,,
\en
and where $\chi$ satisfies the Schr\"odinger equation \eqref{1}. But instead of equation \eqref{2}, the second particle now satisfies 
\be
m_2 {\ddot {\bf X}}_2(t) =   -   {\boldsymbol \nabla}_2   V({\bf X}_1(t),{\bf x}_2)  \big|_{{\bf x}_2={\bf X}_2(t)} \,,
\label{4}
\en
where the force depends on the position of the first particle. So in this approximation the second particle is not acted upon by some average force, but rather by the actual particle of the quantum system. This approximation is therefore expected to yield a better approach than the usual approach, in the sense that it yields predictions closer to those predicted by full quantum theory, especially in the case where the wave function evolves into a superposition of non-overlapping packets. This is indeed confirmed by a number of studies, as we will discuss below.

Let us first mention some properties of this approximation and compare them to the usual approach. In the mean field approach, the specification of an initial wave function $\chi({\bf x}_1,t_0)$, an initial position ${\bf X}_2(t_0)$ and velocity ${\dot {\bf X}}_2(t_0)$ determines a unique solution for the wave function and the trajectory of the classical particle. In the Bohmian approach also the initial position ${\bf X}_1(t_0)$ of the particle of the quantum system needs to be specified in order to uniquely determine a solution. Different initial positions ${\bf X}_1(t_0)$ yield different evolutions for the wave function and the classical particle. This is because the evolution of each of the variables ${\bf X}_1,{\bf X}_2,\chi$ depends on the others. Namely, the evolution of $\chi$ depends on ${\bf X}_2$ via \eqref{1}, whose evolution in turn depends on ${\bf X}_1$ via \eqref{4}, whose evolution in turn depends on $\chi$ via \eqref{3}. (This should be contrasted with the full Bohmian theory, where the wave function acts on the particles, but there is no back-reaction from the particles onto the wave function.) 

The initial configuration ${\bf X}_1(t_0)$ should be considered random with distribution $|\chi({\bf x}_1,t_0)|^2$. However, this does not imply that ${\bf X}_1(t)$ is random with distribution $|\chi({\bf x}_1,t)|^2$ for later times $t$. It is not even clear what the latter statement should mean, since different initial positions ${\bf X}_1(t_0)$ lead to different wave function evolution; so which wave function should $\chi({\bf x}_1,t)$ be? 

This semi-classical approximation has been applied to a number of systems. Prezhdo and Brookby studied the case of a light particle scattering off a heavy particle \cite{prezhdo01}. They considered the scattering probability over time and found that the Bohmian semi-classical approximation was in better agreement with the exact quantum mechanical prediction than the usual approximation. The Bohmian semi-classical approximation gives probability one for the scattering to have happened after some time, in agreement with the exact result, whereas the probability predicted by the usual approach does not reach one. The reported reason for the better results is that the wave function of the quantum particle evolves into a superposition of non-overlapping packets, which yields bad results for the usual approach (since the force on the classical particle contains contributions from both packets), but not for the Bohmian approach. These results were confirmed and further expanded by Gindensperger {\em et al.}\ \cite{gindensperger02a}. Other examples have been considered in \cite{gindensperger00,gindensperger02b,meier04}. In those cases, the Bohmian semi-classical approximation gave very good agreement with the exact quantum or experimental results. It was always either better or comparable to the usual approach. These results give good hope that the Bohmian semi-classical approximation will also give better results than the usual approximation in other domains such as quantum gravity.

\subsection{Derivation of the Bohmian semi-classical approximation}
The Bohmian semi-classical approach can easily be derived from the full Bohmian theory.{\footnote{The derivation is very close to the one followed by Gindensperger {\em et al.}\ \cite{gindensperger00}. A difference is that they also let the wave function of the quantum system depend parametrically on the position of the classical particle. This leads to a quantum force term in the equation \eqref{4} for particle 2. However, this does not seem to lead to a useful set of equations. In particular, they can not be numerically integrated by simply specifying the initial wave function and particle positions. In any case, Gindensperger {\em et al.}\ drop this quantum force when considering examples \cite{gindensperger00,gindensperger02a,gindensperger02b}, so that the resulting equations correspond to the ones presented above.}} Consider a system of two particles. In the Bohmian description of this system, we have a wave function $\psi({\bf x}_1,{\bf x}_2,t)$ and positions ${\bf X}_1(t),{\bf X}_2(t)$, which respectively satisfy the Schr\"odinger equation
\be
\ii \pa_t \psi = \left[ -\frac{1}{2m_1} \nabla^2_1 - \frac{1}{2m_2} \nabla^2_2 + V({\bf x}_1,{\bf x}_2) \right] \psi 
\label{5}
\en
and the guidance equations
\be
{\dot {\bf X}}_1(t) = {\bf v}^\psi_1({\bf X}_1(t),{\bf X}_2(t),t) \,, \qquad {\dot {\bf X}}_2(t) = {\bf v}^\psi_2({\bf X}_1(t),{\bf X}_2(t),t)\,.
\label{6}
\en
The conditional wave function $\chi({\bf x}_1,t) = \psi({\bf x}_1,{\bf X}_2(t),t)$ for particle 1 satisfies the equation
\be
\ii \pa_t \chi({\bf x}_1,t) =  \left( - \frac{\nabla^2_1}{2m_1}  + V({\bf x}_1,{\bf X}_2(t)) \right) \chi({\bf x}_1,t) + I({\bf x}_1,t) \,,
\label{7}
\en
where
\be
I({\bf x}_1,t) =  \left(- \frac{\nabla^2_2}{2m_2}  \psi({\bf x}_1,{\bf x}_2,t) \right) \Bigg|_{{\bf x}_2={\bf X}_2(t)} +  \ii {\boldsymbol \nabla}_2 \psi({\bf x}_1,{\bf x}_2,t)\Big|_{{\bf x}_2={\bf X}_2(t)} \cdot {\bf v}^\psi_2({\bf X}_1(t),{\bf X}_2(t),t) \,.
\label{8}
\en
So in case $I$ is negligible in \eqref{7}, up to a time-dependent factor times $\chi$,{\footnote{If $I$ contains a term of the form $f(t)\chi$, then it can be eliminated by changing the phase of $\chi$ by a time-dependent term.\label{timedependent}}} we are led to the Schr\"odinger equation \eqref{1}. This will for example be the case if $m_2$ is much larger than $m_1$ ($I$ is inversely proportional to $m_2$) and if the wave function slowly varies as a function of ${\bf x}_2$. We also have that
\be
m_2 {\ddot {\bf X}}_2(t) =   - {\boldsymbol \nabla}_2   \left[ V({\bf X}_1(t),{\bf x}_2) + Q^\psi({\bf X}_1(t),{\bf x}_2,t) \right]  \Bigg|_{{\bf x}_2={\bf X}_2(t)}\,,
\en
with $Q^\psi$ the quantum potential. We obtain the classical equation \eqref{4}, if the quantum force is negligible compared to the classical force. 

In this way we obtain the equations for a semi-classical formulation. In addition, we also have the conditions under which they will be valid. For other quantum theories, such as quantum gravity, we can follow a similar path to find a Bohmian semi-classical approximation.

\section{Scalar electrodynamics}\label{sqed}
We consider scalar electrodynamics to illustrate the issues with developing a Bohmian semi-classical approximation for a gauge theory. There are various {\em  equivalent} ways of formulating the Bohmian approach \cite{struyve15}. These formulations can either be found by considering different gauges or by working with different choices of gauge-independent variables. Here, we will consider two examples of gauges, namely the temporal gauge, which is an incomplete gauge fixing, and the Coulomb gauge, which completely fixes the gauge symmetry. Using the former gauge, we are not immediately led to a semi-classical approximation, due to the remaining gauge freedom, while we {\em are}, using the latter gauge.

In classical scalar electrodynamics, the equations of motion for the scalar field $\phi$ and the vector potential $A^\mu=(A_0,{\bf A})$ are 
\begin{equation}
D_{\mu}D^{\mu}\phi + m^2 \phi = 0\,, \qquad \partial_{\mu} F^{\mu \nu}= j^{\nu} \,,
\label{sq.2}
\end{equation}
where $D_{\mu}= \partial_{\mu} + \ii e A_{\mu}$ is the covariant derivative, $F^{\mu \nu} = \pa^\mu A^\nu - \pa^\nu A^\mu$ the electromagnetic field tensor and
\be
j^{\nu} = \ii e \left(\phi^*D^{\nu}\phi - \phi D^{\nu*} \phi^* \right) 
\label{sq.3}
\en
is the charge current. The theory has a local gauge symmetry
\begin{equation}
\phi \to e^{\ii e \alpha} \phi \,,  \quad  A^{\mu} \to   A^\mu -  \partial^\mu  \alpha\,. 
\label{sq.4}
\end{equation}

One possible choice of gauge is the temporal gauge $A_0 = 0$. It does not completely fix the gauge; there is still a residual gauge symmetry given by the time-independent transformations
\be
\phi \to \ee^{\ii e \theta} \phi \,, \qquad {\bf A} \to {\bf A} + \bn \theta \,,
\label{tg.001}
\en
with ${\dot \theta} = 0$. Quantization in this gauge leads to the following functional Schr\"o\-din\-ger equation for $\Psi(\phi,{\bf A},t)$ \cite{kiefer92}:{\footnote{The wave functional should be understood as a functional of the real and imaginary part of $\phi$. In addition, writing $\phi = (\phi_r + \ii \phi_i)/{\sqrt 2}$, we have that the functional derivatives are given by $\delta / \delta \phi = (\delta / \delta \phi_r  - \ii \delta / \delta \phi_i)/{\sqrt 2}$ and $\delta / \delta \phi^* = (\delta / \delta \phi_r  + \ii \delta / \delta \phi_i)/{\sqrt 2}$.}}
\be
\ii \pa_t \Psi =   \int d^3 x \bigg(-\frac{ \delta^2  }{ \delta \phi^* \delta \phi} + |{\bf D} \phi|^2  + m^2 |\phi|^2 - \frac{1}{2} \frac{\delta^2}{ \delta {\bf A}^2} + \frac{1}{2} (\bn \times {\bf A})^2  \bigg)\Psi \,,   
\label{tg.01}
\en
together with the constraint
\be
\bn \cdot \frac{\delta \Psi} {  \delta {\bf A} } + \ii e \left(\phi^* \frac{ \delta \Psi }{\delta \phi^*} -  \phi \frac{ \delta \Psi }{ \delta \phi} \right) = 0 \,.
\label{tg.02}
\en
The constraint expresses the fact that the wave functional is invariant under time-independent gauge transformations, i.e., $\Psi(\phi,{\bf A})=\Psi(\ee^{\ii e \theta}\phi,{\bf A} +  \bn \theta)$, with $\theta$ time-independent. The constraint is compatible with the Schr\"odinger equation: if it is satisfied at one time, it is satisfied at all times. 

In the Bohmian approach \cite{valentini92}, there are actual configurations $\phi$ and ${\bf A}$ that satisfy the guidance equations
\be
{\dot \phi} =  \frac{\delta S}{\delta \phi^* } \,, \qquad {\dot {\bf A}} =  \frac{\delta S}{\delta {\bf A}  } \,,
\label{tg.03}
\en
where $\Psi = |\Psi|\ee^{\ii S}$. These equations are invariant under the time-independent gauge transformations \eqref{tg.001} because of the constraint \eqref{tg.02}.

In the framework of standard quantum theory, there is a natural semi-classical approximation that treats the vector potential classically and the scalar field quantum mechanically. The scalar field is described by a wave functional $\chi(\phi,t)$ which satisfies
\be
\ii \pa_t \chi =   \int d^3 x \bigg(-\frac{ \delta^2  }{ \delta \phi^* \delta \phi} + |{\bf D} \phi|^2  + m^2 |\phi|^2 \bigg)\chi    
\label{tg.0201}
\en
and the electromagnetic field satisfies Maxwell's equations (with $A_0=0$)
\be
\pa_\mu F^{\mu \nu} = \langle \chi | {\widehat j}^\nu | \chi \rangle  \,,
\label{tg.0202}
\en
where
\be
\langle \chi | {\widehat j}^0 | \chi \rangle = \int {\mathcal D} \phi \Psi^* {\mathcal C} \Psi =  e \int {\mathcal D} \phi \Psi^* \left(\phi^* \frac{ \delta \Psi }{\delta \phi^*} -  \phi \frac{ \delta \Psi }{ \delta \phi} \right) \,,
\nonumber
\en
\be
\langle \chi |{\widehat {\bf j}} | \chi \rangle = \ii  e \int {\mathcal D} \phi |\Psi|^2  \left(\phi {\bf D}^*\phi^* - \phi^* {\bf D} \phi\right)\,,
\label{tg.0203}
\en
with
\be
{\mathcal C} ({\bf x}) = e \left(\phi^*({\bf x}) \frac{ \delta  }{\delta \phi^*({\bf x})} -  \phi({\bf x}) \frac{ \delta }{ \delta \phi({\bf x})} \right)
\en
the charge density operator in the functional Schr\"odinger picture. This theory is consistent since $\pa_\mu \langle \chi | {\widehat j}^\mu | \chi \rangle = 0$, as a consequence of the Schr\"odinger equation \eqref{tg.0201}. 

A natural guess for a Bohmian semi-classical approximation similar to the usual one is the following (and can be obtained from the full Bohmian approach by considering the conditional wave function $\chi(\phi,t) = \Psi(\phi,{\bf A}(t),t)$). An actual field $\phi$ is introduced that satisfies ${\dot \phi} =  \delta S/\delta \phi^*$, where the wave functional satisfies \eqref{tg.0201}, and Maxwell's equations read $\pa_\mu F^{\mu \nu} = j^\nu$, where $j^\mu$ is the classical expression for the charge current. However, the second order equation for the Bohmian field is
\be
{\ddot \phi} - D^2 \phi + m^2 \phi = - \frac{\delta Q^\chi}{\delta \phi^*} \,, 
\label{tg.04}
\en
where $Q^\chi = - \frac{1}{|\chi|} \int d^3 x \left( \frac{\delta^2 |\chi|}{\delta \phi^* \delta \phi}\right)$. As a consequence, $\pa_\mu j^\mu = - \ii {\mathcal C} Q^\chi $ and hence Maxwell's equations imply that ${\mathcal C} Q^\chi =0$ or $Q^\chi=Q^\chi(|\phi|^2)$. This is a constraint on the wave functional that was absent in the usual semi-classical theory. It also seems to be a rather strong condition. It will for example be satisfied if the scalar field evolves classically (i.e., when the right-hand side of \eqref{tg.04} is zero) but it is unclear whether there are other solutions.

So the conclusion seems to be that if we assume ${\bf A}$ classical, then $\phi$ should also behave classically. This is not surprising since the gauge symmetry implies that the physical (i.e., gauge invariant) degrees of freedom are some combination of the fields ${\bf A}$ and $\phi$. So one can not just assume ${\bf A}$ classical and keep $\phi$ fully quantum.

In \cite{struyve15}, we showed that the problem disappears if we eliminate the gauge freedom, for example by using a gauge which completely fixes the gauge freedom. We discussed in detail the Coulomb and the unitary gauge and showed that a semi-classical approximation can easily be obtained by considering either the scalar or electromagnetic field classically. 

Let us consider the Coulomb gauge  ${\boldsymbol \nabla} \cdot {\bf A}=0$ here. In the full Bohmian approach \cite{struyve10,struyve15} we have that there are actual fields{\footnote{We have used the decomposition ${\bf A}={\bf A}^T + {\bf A}^L$, where ${\bf A}^T$ and ${\bf A}^L$ are respectively the transverse and longitudinal part of the vector potential. The Coulomb gauge then corresponds to ${\bf A}^L = 0$.}} $\phi$ and ${\bf A}^T$ that are guided by a wave functional $\Psi(\phi,{\bf A}^T,t)$ which satisfies the functional Schr\"odinger equation
\be
\ii \pa_t \Psi =   \int d^3 x \bigg(-\frac{ \delta^2  }{ \delta \phi^* \delta \phi} + | (\bn - \ii e {\bf A}^T) \phi|^2  + m^2 |\phi|^2     
- \frac{1}{2} {\mathcal C} \frac{1}{\nabla^2} {\mathcal C}  
 - \frac{1}{2} \frac{\delta^2}{ \delta {\bf A}^{T2}} + \frac{1}{2}(\bn \times {\bf A}^T)^2 \bigg)\Psi \,.
\label{sq.8}
\en
 The first three terms in the Hamiltonian correspond to the Hamiltonian of a scalar field minimally coupled to a transverse vector potential. The fourth term corresponds to the Coulomb potential and the remaining terms to the Hamiltonian of a free electromagnetic field. The guidance equations are
\be
{\dot\phi } =  \frac{\delta S}{\delta \phi^* }  - e \phi \frac{1}{\nabla^2}{\mathcal C} S \,,\qquad \dot {\bf A}^T =  \frac{\delta S}{\delta {\bf A}^T  } \,.
\en
Defining 
\be
A_0 = - \ii \frac{1}{\nabla^2}{\mathcal C} S \,,
\en
we can rewrite the guidance equation for the scalar field as
\be
D_0 \phi = \frac{\delta S}{\delta \phi^*} \,.
\en
The definition of $A_0$ was motivated by analogy with the classical equations of motion \cite{struyve15}.

While this Bohmian approach is equivalent to the one in the temporal gauge \cite{struyve15}, it naturally leads to the following semi-classical approximation (by considering the conditional wave function for the scalar field). The wave functional $\chi(\phi,t)$ satisfies
\be
\ii \pa_t \chi =   \int d^3 x \bigg(-\frac{ \delta^2  }{ \delta \phi^* \delta \phi} +  | (\bn - \ii e {\bf A}^T) \phi|^2 + m^2 |\phi|^2  - \frac{1}{2} {\mathcal C} \frac{1}{\nabla^2} {\mathcal C} \bigg)\chi    
\label{sq.11}
\en
and guides the actual scalar field through
\be
D_0 \phi = \frac{\delta S}{\delta \phi^*}  \,,
\label{sq.12}
\en
where $A_0$ is defined as before and with $S$ now the phase of $\chi$. The vector potential $A^\mu = (A_0, {\bf A}^T)$ satisfies Maxwell's equations
\be
\pa_\mu F^{\mu \nu} = j^\nu + j^\nu_Q\,,
\label{sq.13}
\en
where $j^\nu_Q = (0,{\bf j}_Q)$ is an additional ``quantum'' current, with
\be
{\bf j}_Q=  \ii {\boldsymbol \nabla} \frac{1}{\nabla^2}{\mathcal C} Q^\chi  
\en
and 
\be
Q^\chi = - \frac{1}{|\chi|} \int d^3 x \left( \frac{\delta^2 |\chi|}{\delta \phi^* \delta \phi} + \frac{1}{2}{\mathcal C} \frac{1}{\nabla^2} {\mathcal C} |\chi| \right)
\en
the quantum potential. These equations are consistent in the sense that $\pa_\mu (j^\mu + j^\mu_Q) = 0$, as a consequence of the second order equation 
\be
D_\mu D^\mu \phi - m^2 \phi = - \frac{\delta Q^\chi}{\delta \phi^*} \,,
\label{sq.14}
\en
which follows from taking the time derivative of \eqref{sq.12}.

\section{Quantum gravity: mini-superspace model}\label{mini}
The structure of the Bohmian approach to canonical gravity (outlined in section \ref{qg}) is similar to that of scalar electrodynamics in the temporal gauge. Namely, in both cases there is a constraint on the wave functional which expresses invariance under infinitesimal gauge transformations: spatial diffeomorphisms in the former case and phase transformations in the latter case. In both cases the gauge invariance seems to be the source of the problem in formulating a consistent Bohmian semi-classical approximation. In the case of quantum electrodynamics the problem was overcome by gauge fixing. Presumably one can find a similar solution in quantum gravity. However, finding a suitable gauge is a notoriously hard problem in this case. We can however consider a mini-superspace model which is a symmetry-reduced approach to quantum gravity where homogeneity and isotropy are assumed. In this case, the spatial diffeomorphism invariance is eliminated and we can straightforwardly develop a Bohmian semi-classical approximation, as we will now show.

In the classical mini-superspace model, the universe is described by the Friedmann-Lema\^itre-Robertson-Walker (FLRW) metric 
\be
\ud s^2 =  N(t)^2 \ud t^2 - a(t)^2 \ud \Omega^2_3 \,,
\label{ms.01}
\en
where $N$ is the lapse function, $a=\ee^{\al}$ is the scale factor{\footnote{The reason for introducing the variable $\al$ is that it is unbounded, unlike the scale factor, which satisfies $a>0$.}}  and $\ud \Omega^2_3$ is the metric on 3-space with constant curvature $k$. Assuming matter that is described by a homogeneous scalar field $\phi$, the equations of motion are \cite{halliwell91b,vink92,struyve15}:
\begin{align}
& \frac{1}{2} {\dot \al}^2 = \frac{1}{\ka}\left(\frac{1}{2} {\dot \phi}^2 +V_M\right) + V_G \,,\label{ms.031}\\
& \ddot \phi + 3 \dot \al \dot \phi + \pa_\phi V_M = 0 \,,
\label{ms.04}
\end{align}
where the gauge $N=1$ is choosen,{\footnote{The theory is time-reparamaterization invariant. Solutions that differ only by a time-reparameterization are considered physically equivalent. Choosing the gauge $N=1$ corresponds to a particular time-parameterization.}}  $\ka = 3/4\pi G$,
\be
V_G = - \frac{1}{2} k \ee^{-2\al} + \frac{1}{6} \Lambda 
\label{ms.03}
\en
is the gravitational potential, with $\Lambda$ the cosmological constant, and $V_M$ is the potential for the matter field. 

Canonical quantization of the classical theory yields the Wheeler-DeWitt equation:
\be
({\widehat H}_G + {\widehat H}_M) \psi = 0 \,,
\label{ms.07}
\en
where
\be
{\widehat H}_G = \frac{1}{2\ka \ee^{3\al}} \pa^2_\al + \ka \ee^{3\al}V_G \,, \qquad {\widehat H}_M = - \frac{1}{2\ee^{3\al}} \pa^2_\phi + \ee^{3\al}V_M \,.
\label{ms.08}
\en
In the corresponding Bohmian approach \cite{vink92}, there is an actual FLRW metric of the form \eqref{ms.01} and scalar field, whose time evolutions are determined by the guidance equations
\be
\dot \al = -  \frac{N}{\ka\ee^{3\al}} \pa_\al S \,, \qquad \dot \phi= \frac{N}{\ee^{3\al}} \pa_\phi S \,, 
\label{ms.09}
\en 
where $N$ is an arbitrary lapse function.{\footnote{Just as the classical theory, the Bohmian approach is time-reparameterization invariant. This is a special feature of mini-superspace models \cite{acacio98,falciano01}. As mentioned before, for the usual formulation of the Bohmian dynamics for the Wheeler-DeWitt theory of quantum gravity, a particular space-like foliation of space-time or, equivalently, a particular choice of ``initial'' space-like hypersurface and lapse function, needs to be introduced. Different foliations (or lapse functions) yield different Bohmian theories.}} In the gauge $N=1$, these equations imply 
\begin{align}
& \frac{1}{2} {\dot \al}^2 = \frac{1}{\ka}\left(\frac{1}{2} {\dot \phi}^2 +V_M + Q^\psi_M \right) + V_G + Q^\psi_G \,,\label{ms.09.1}\\
& \ddot \phi + 3 \dot \al \dot \phi + \pa_\phi (V_M + Q^\psi_M + \ka Q^\psi_G) = 0 \,,
\label{ms.09.2}
\end{align}
where 
\be
Q^\psi_G = \frac{1}{2\ka^2 \ee^{6\al} } \frac{\pa^2_\al |\psi|}{|\psi|} \,, \qquad Q^\psi_M = - \frac{1}{2 \ee^{6\al} } \frac{\pa^2_\phi |\psi|}{|\psi|}
 \,.
\label{ms.10}
\en

We will now look for a semi-classical approximation where the scale factor behaves approximately classical. In order to do so, we assume again the gauge $N=1$ and we consider the conditional wave function $\chi(\phi,t) = \psi (\phi,\al(t))$, given a set of trajectories $(\al(t),\phi(t))$. Using 
\be
\pa_t \chi (\phi,t) = \pa_\al \psi(\phi,\al) \big|_{\al = \al(t)} \dot \al(t) \,,
\label{ms.11}
\en
we can write
\be
\ii \pa_t \chi =  {\widehat H}_M \chi + I \,,
\label{ms.12}
\en
where{\footnote{To obtain this equation, note that $\pa^2_\al \psi = [ (\pa_\al S)^2 + \ii\pa^2_\al S  + \pa^2_\al |\psi|/|\psi|]\psi +2\ii \pa_\al S \pa_\al \psi$, so that $\pa^2_\al \psi|_{\al = \al(t)} = [ (\pa_\al S)^2 + \ii\pa^2_\al  + \pa^2_\al |\psi|/|\psi|]|_{\al = \al(t)}\chi +2\ii  \pa_\al S \pa_t \chi / \dot \al$. Using this equation together with \eqref{ms.07} we obtain \eqref{ms.12}.}}
\be
I = \frac{1}{\dot \al}\ii \pa_t \chi \left( \dot \al + \frac{1}{ \ka\ee^{3\al}} \pa_\al S \big|_{\al(t)}\right) + \frac{1}{2\ka\ee^{3\al}}\left[ (\pa_\al S)^2 + \ii\pa^2_\al S\right]\Big|_{\al(t)}\chi  +\ka \ee^{3\al} (V_G + Q^\psi_G)  \Big|_{\al = \al(t)} \chi \,.
\label{ms.13}
\en
When $I$ is negligible (up to a real time-dependent function times $\chi$), \eqref{ms.12} becomes the Schr\"odinger equation for a homogeneous matter field in an external FLRW metric. We can further assume the quantum potential $Q^\psi_G$ to be negligible compared to other terms in eq.\ \eqref{ms.09.1}. As such, we are led to the semi-classical theory:
\begin{align}
& \ii \pa_t \chi =  {\widehat H}_M \chi \,,\label{ms.14}\\
& \dot \phi=   \frac{1}{\ee^{3\al}} \pa_\phi S \,,\label{ms.15}\\
& \frac{1}{2} {\dot \al}^2 = \frac{1}{\ka}\left(\frac{1}{2} {\dot \phi}^2 +V_M +Q^\chi_M\right) + V_G \equiv  - \frac{1}{\ka\ee^{3\al}} \pa_t S  + V_G  \,. \label{ms.16}
\end{align}

Let us now consider when the term $I$ will be negligible. The quantity in brackets in the first term would be zero when evaluated for the actual trajectory $\phi(t)$ (because of the guidance equation for $\al$). As such, the first term will be negligible if the actual scale factor evolves approximately independently of the scalar field. The second term will be negligible if $S$ varies slowly with respect to $\al$ or if the term in square brackets is approximately independent of $\phi$. In the latter case, the second term becomes a time-dependent function times $\chi$, which can be eliminated by changing the phase of $\chi$. Similarly, if $Q^\psi_G \ll V_G$ then the third term also becomes a time-dependent function times $\chi$. 

In the usual semi-classical approximation, one has \eqref{ms.14} and
\be
\frac{1}{2} {\dot \al}^2 = \frac{1}{\ka \ee^{3\al}}\langle \chi| {\widehat H}_M| \chi \rangle + V_G \,, 
\en
with $\chi$ normalized to one. These equations follow from \eqref{0.001} and \eqref{0.002}. In \cite{struyve15} an example is worked out for which the Bohmian semi-classical approximation gives better results than this approximation. (Note that Vink himself, in his seminal paper on applying the Bohmian approach to quantum gravity, considers a derivation of the usual semi-classical approximation, rather than the Bohmian one. But he hinted on the Bohmian semi-classical approximation in \cite{kowalski-glikman90}.)

\section{Conclusion}
We have shown how semi-classical approximations can be developed using Bohmian mechanics. We have obtained these approximations from the full Bohmian theory by assuming certain degrees of freedom to evolve approximately classically. This was illustrated for non-relativistic systems. If there is a gauge symmetry, like in electrodynamics or gravity, then extra care is required in order to obtain a consistent semi-classical theory. By eliminating the gauge symmetry, either by imposing a gauge or by working with gauge-independent degrees of freedom, we were able to find a semi-classical approximation in the case of scalar quantum electrodynamics. For quantum gravity, eliminating the gauge symmetry (more precisely the spatial diffeomorphism invariance) is notoriously hard. We have only considered the simplified mini-superspace approach to quantum gravity, which describes an isotropic and homogeneous universe, and where the diffeomorphism invariance is explicitly eliminated. More general cases in quantum gravity still need to be studied. For example, for the case of inflation theory, where one usually considers fluctuations around an isotropic and homogeneous universe, it should not be too difficult to develop a Bohmian semi-classical approximation. 

Apart from possible applications in quantum cosmology, such as inflation theory, it might also be interesting to consider potential applications in quantum electrodynamics or quantum optics. In particular, since the results may be compared to the predictions of full quantum theory, this could may give us a handle on where to expect better results for the Bohmian semi-classical approximation compared to the usual one in the case of quantum gravity where the full quantum theory is not known. That is, it might give us better insight in which effects are truly quantum and which effect are merely artifacts of the approximation.

Further developments may include higher order corrections to the semi-classical approximation. One way of doing this might be by following the ideas presented in \cite{norsen10,norsen15}. As explained there, one might introduce extra wave functions for a subsystem in addition to the conditional wave function. These wave functions interact with each other and the Bohmian configurations. By including more of those wave functions one presumably obtains better approximations to the full quantum result.

Finally, although we regard the Bohmian semi-classical approximation for quantum gravity as an approximation to some deeper quantum theory for gravity, one could also entertain the possibility that it is a fundamental theory on its own. At least, there is presumably as yet no experimental evidence against it.

\section{Acknowledgments}
This work is supported by the Actions de Recherches Concert\'ees (ARC) of the Belgium Wallonia-Brussels Federation under contract No.\ 12-17/02.


\end{document}